\documentclass[twocolumn,aps,prl,showpacs,preprintnumbers,amssymb]{revtex4-1}

\usepackage{graphicx}
\usepackage{dcolumn}
\usepackage{bm}
\usepackage{epsf}
\usepackage{natbib}
\usepackage{hyperref}
\usepackage{amsmath}
\usepackage{multirow}

\begin{document}
\newcommand{\cm}{\ \mbox{cm}^{-1} }

\providecommand{\mean}[1]{\ensuremath{\left \langle #1 \right \rangle}}
\providecommand{\abs}[1]{\ensuremath{\left | #1 \right |}}

\title{The inverted pendulum, interface phonons and optic Tamm states}

\author{Nicolas Combe}
\affiliation{Centre d'Elaboration de Mat\'eriaux et d'Etudes
Structurales, CNRS UPR 8011, 29 rue J. Marvig, BP 94347, 31055 Toulouse cedex
4, France}
\affiliation{Universit\'e de Toulouse ; UPS ; F-31055 Toulouse, France}

\providecommand{\ket}[1]{\ensuremath{\left | #1 \right \rangle}}
\providecommand{\bra}[1]{\ensuremath{\left \langle #1 \right |}}
\providecommand{\braket}[2]{\ensuremath{\left \langle   #1  |  #2 \right \rangle}}
\date{\today}

\begin{abstract}
 The propagation of waves in periodic media is related to the parametric oscillators. We transpose the possibility that a parametric pendulum oscillates in the vicinity of its  unstable equilibrium positions to the case of waves in lossless unidimensional periodic media. This concept  formally applies to any kind of wave. We apply and develop it to the case of phonons in realizable structures and evidence unconventional types of phonons.  Discussing the case of electromagnetic waves,  we show that our concept is related to optic Tamm states one but extends it to periodic Optic Tamm state. 
\end{abstract}


\maketitle


Waves in periodic media have focused the interest of many scientists and have found numerous applications in different branches of physics: 
electromagnetic (EM) waves in photonic crystals~\cite{Istrate2006}, phonon (elastic wave)  in crystalline solids or in phononic crystals~\cite{Boudouti2009}, electron wave functions in crystalline solids~\cite{ashcroft} or electronic super-lattices~\cite{Steslicka2002}.
The periodicity of the media yields the existence of band gaps (BG) in which the amplitude of the wave exponentially varies, hence corresponding to unphysical states in infinite media. Outside of these gaps, waves are spatially periodic. In the presence of localized defects, localized modes, including surface or interface modes have been evidenced.

\begin{figure}[h!]
\begin{center}
a)\includegraphics[width=2.5cm]{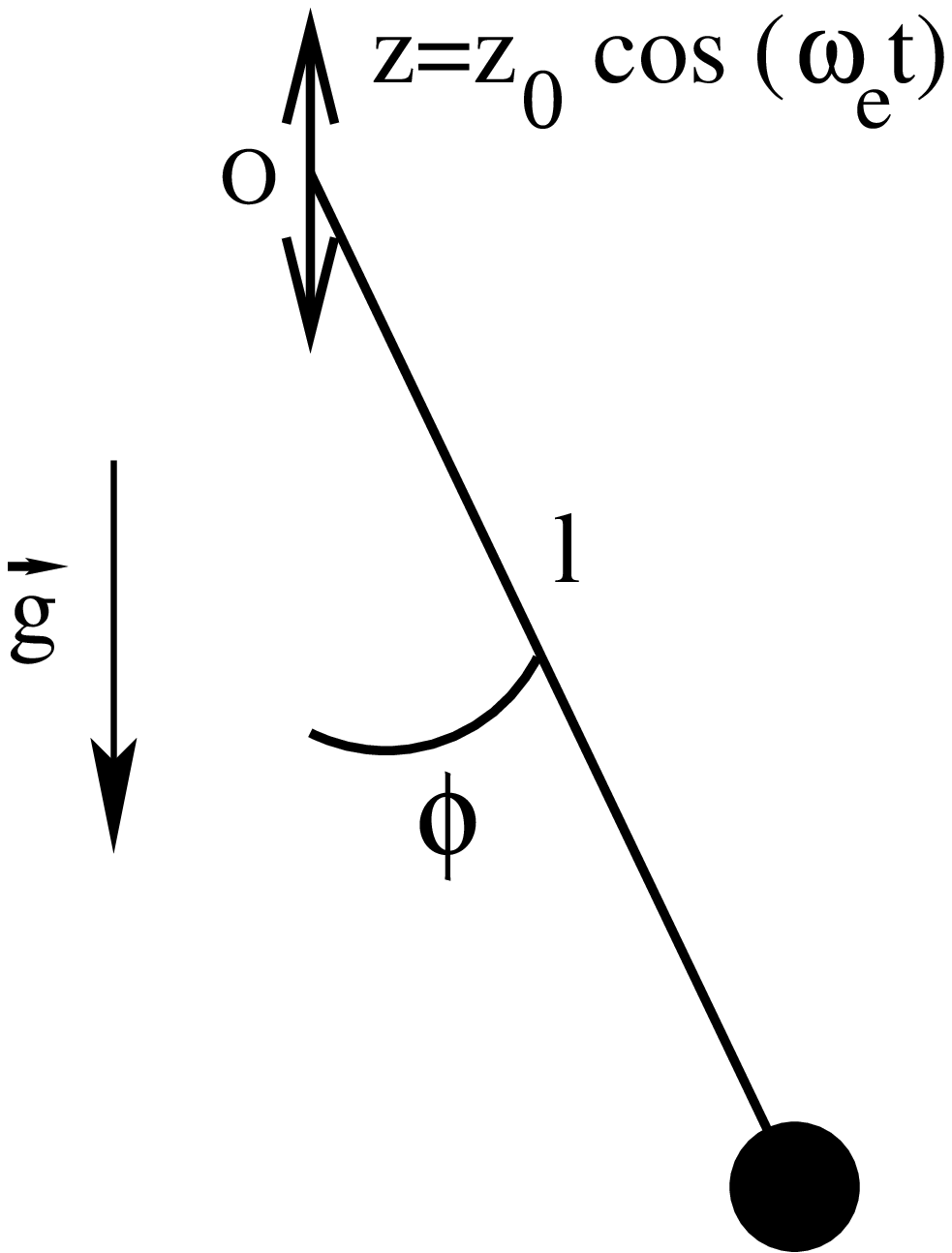}
b)\includegraphics[width=5cm]{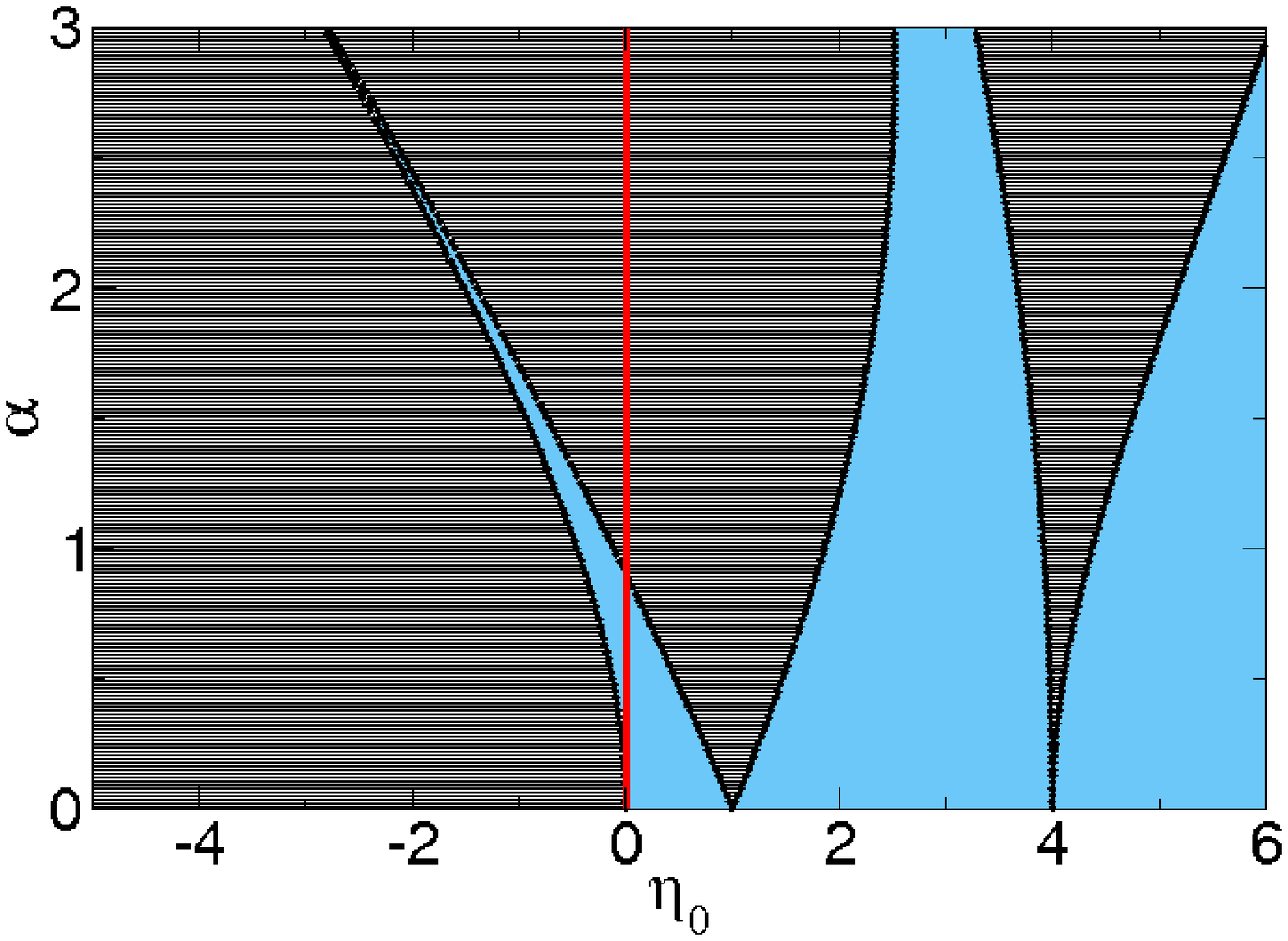}
\caption{a) Sketch of a  parametric pendulum b) Phase diagram of Eq.~\eqref{eq_mat_ssdim}. Periodic and unbounded solutions are in blue and dashed regions.  A vertical red line points up  $\eta_0=0$.} 
\label{fig1}
\end{center}
\end{figure}

As exploited recently~\cite{Combe2009a},  the physic of  the parametric oscillator is equivalent to that of the propagations of  waves in a lossless unidimensional infinite periodic (LUIP) medium. In this manuscript, we transpose the striking possibility for an oscillator to oscillate in the vicinity of an unstable equilibrium position using a parametric excitation (in the inverted pendulum experiment, for instance)  to the case of waves in LUIP media. We evidence unconventional types of waves. This concept is  general and can formally apply to any kind of waves. \\
Before exposing and applying this concept, we first explicit the equivalence between  the parametric oscillator and  the propagations of  waves in LUIP media.\\ 
Main features of the parametric oscillator infer from the fixed point stability of a parametrically and sinusoidally excited pendulum (for instance, see Fig.~\ref{fig1}a) governed by the Mathieu equation: 
\begin{eqnarray}
\frac{d^2 \phi}{d\tilde{t}^2} + [ \eta_0 + 2 \alpha \cos(2\tilde{t})] \phi =0 \label{eq_mat_ssdim}
\end{eqnarray}
with $\tilde{t} = {\omega_et}/{2} $, $\eta_0= \pm { 4 \omega_0^2}/{\omega_e^2}$ and $\alpha = {-2 z_0}/{l}$, $\omega_0^2={g}/{l}$.
where $\phi$, $l$, $g$, $\omega_e$ and $z_0$ are respectively the angle and length of the pendulum, the standard gravity and the excitation frequency and amplitude.
$\eta_0$ is positive around the fixed point $\vec{\Phi}=(\phi,\frac{d \phi}{d\tilde{t}})= (0,0)$ and negative around $\vec{\Phi}= (\pi,0)$. 
Depending on $\eta_0$ and $\alpha$, solutions of Eq.~\eqref{eq_mat_ssdim} are either periodic or unbounded. Unbounded solutions, corresponding to the parametric resonance oscillate with an exponentially varying amplitude.  Fig.~\ref{fig1}b reports the phase diagram of Eq.~\eqref{eq_mat_ssdim}: using the Floquet theory, it deduces from the eigenvalues of ${\bf R}_0^{\pi}$ with ${\bf R}_{\tilde{t}_0}^{\tilde{t}}$ the propagator of Eq.~\eqref{eq_mat_ssdim}: $\vec{\Phi}(\tilde{t}) = {\bf R}_{\tilde{t}_0}^{\tilde{t}} \vec{\Phi}(\tilde{t}_0)$.
Remarkably, Fig.~\ref{fig1}b evidences some periodic solutions in the $\eta_0<0$ region, corresponding to the inverted pendulum experiment~\cite{Kapitza1965}.

We consider now the propagation of waves in a LUIP medium and focus on EM and elastic waves to reveal the equivalence between the parametric oscillator and the propagation of waves in a LUIP medium.\\
The time Fourier transform  $\vec{E}(z,\omega)$ of the electric field  $\vec{\mathfrak{E}}(z,t) = \vec{E}(z,\omega) e^{i \omega t}$ of an EM wave in a LUIP medium with normal incidence is solution of: 
  \begin{equation}
  \frac{d^2 \vec{E}}{dz^2}(z,\omega)  + p(z) \vec{E}(z,\omega) =\vec{0} \label{EM}
  \end{equation}
   Where $p(z) = \frac{ \epsilon(z)\omega^2}{c^2}$ and $\epsilon(z) $ and $c$ are the relative dielectric permittivity and the light speed in vacuum. \\
    For a longitudinal elastic wave with normal incidence, the time Fourier transform of the projected wave equation of the phonon displacement field $\vec{U}(z,t)= \mathfrak{U}(z,\omega) e^{i \omega t} \vec{z}$ writes using the linear elasticity theory:
\begin{equation}
{C}(z)  \frac{d^2 \mathfrak{U}}{dz^2} (z,\omega) +  \frac{d C}{dz}(z)  \frac{d \mathfrak{U}}{dz}(z,\omega)  + \rho (z) \omega^2 \mathfrak{U}(z,\omega)  =0 \label{eq_wave2}
\end{equation}
where $C$ and $\rho(z)$ are the elastic coefficient $c_{zzzz}$ and the volumic mass.  
Actually, both equations  Eq.~\eqref{EM} and Eq.~\eqref{eq_wave2} describe the same physic. A mathematical transformation of Eq.~\eqref{eq_wave2}
evidences this point: setting $\mathfrak{U}(z,\omega) =  Q(z,\omega) u(z,\omega)$ with $Q(z,\omega)$ satisfying $2 C \frac{dQ}{dz} +\frac{dC}{dz} Q  =0$,  $u(z,\omega)$  is solution of:  
 \begin{equation}
\frac{d^2 u}{dz^2}(z,\omega)  + p(z) u(z,\omega)   =0 \label{eq}
\end{equation}
Where $p(z) = \frac{\rho \omega^2}{C} + \frac{1}{2}\frac{d^2C}{dz^2} - \frac{ 1 }{4C^2} [\frac{dC}{dz}]^2 $ is a generally positive quantity.
Eq.~\eqref{EM} and~\eqref{eq} are actually equivalent.\\
In a LUIP medium,  the function $p(z)$ is real and periodic (period $\lambda$) and Eq.~\eqref{EM}(Eq.~\eqref{eq}) is a  Hill equation, the equation of a parametric oscillator.  Considering a sinusoidal variation of $p(z)$ or limiting the Fourier series of $p(z) \approx p_0 + p_1 \cos(\frac{2 \pi z}{\lambda})$, we recover the Mathieu equation
\begin{eqnarray}
\frac{d^2 \phi}{d\tilde{z}^2} + [ \eta_0 + 2 \alpha \cos(2\tilde{z})] \phi =0 \label{eq_mat_ssdim2}
\end{eqnarray}
Where $\eta_0 = \frac{p_0 \lambda^2}{\pi^2}$, $\alpha = \frac{p_1 \lambda^2}{2 \pi^2}$, $\tilde{z}= \frac{\pi z}{\lambda}$, and $\phi$ designs either $\vec{E}(z,\omega)$ from Eq.~\eqref{EM} or  $u(z,\omega)$ from Eq.~\eqref{eq}. \\
Eq.~\eqref{eq_mat_ssdim} for the parametric oscillator  and~\eqref{eq_mat_ssdim2} for waves are equivalent: they have thus the same solutions~\footnote{Actually, Hill equations solutions are qualitatively identical to Mathieu equation solutions.}. 
Periodic solutions of  Eq.~\eqref{eq_mat_ssdim2} correspond to periodic waves whereas unbounded ones, to waves in the BGs. 

In this study, we address the following questions:
Do the periodic solutions in the phase space $\eta_0<0$ of Fig.~\ref{fig1}b exist for waves in real, eventually advanced materials? if yes to what kind of waves do they correspond?

 $p(z)$ in Eq.~\eqref{EM} is either positive (noticeably in dielectric media) or negative (noticeably in metals below their plasma frequency). The propagation of EM waves in a LUIP metallic system below the metals plasma frequencies seems a straightforward transposition of the inverted pendulum case: the negative values of the dielectric permittivity  warrants a negative value of  $\eta_0$ in Eq.~\eqref{eq_mat_ssdim2}. Studying the dispersion diagram of a theoretical LUIP metallic layered system, we have evidenced some periodic EM waves that propagate below the metals plasma frequencies. However,  in real metals, $p(z)$ in Eq.~\eqref{EM} is a complex quantity whose imaginary part is related to the absorption.  A parametric excitation cannot affect the dissipation. Investigating bi-metallic structures based on real common pure metals~\cite{Palik1997}, the absorption dominates and screens the effect we wish to evidence.
 
To overcome the absorption drawback of metals, we investigate in the following weakly absorbing materials i.e. dielectric materials and neglect their absorption. The dielectric permittivity in Eq.~\eqref{EM} is thus real and positive i.e. a case equivalent to the propagation of the elastic waves:  $p(z)>0 \ \forall z$ in Eq.~\eqref{eq}. 

Waves in the BGs of a LUIP medium oscillate with an exponentially varying amplitude: this amplitude hence varies as the angle of a free pendulum closed to its unstable equilibrium position.  The amplitude of the EM  or elastic waves in BGs behaves as a wave in a negative dielectric permittivity medium or in a hypothetical imaginary sound speed medium i.e. mimics a negative effective $\eta_0$ in Eq.~\eqref{eq_mat_ssdim2} with $\phi$ the amplitude of the wave.

To model the inverted pendulum stabilization mechanism, we propose to use a medium displaying two periodicities: a first one that would create a BG (in the absence of the second periodicity) and mimic a negative effective $\eta_0$, and a second one that parametrically creates some periodic solutions inside this BG. 
Eq.~\eqref{eq_mat_ssdim2} is formally replaced by the following modified Mathieu equation: 
\begin{equation}
\frac{d \phi}{d\tilde{z}^2} + ( \eta_0 + 2 \alpha \cos(2 \tilde{z}) +   \beta \cos(k_e \tilde{z} + \varphi) )  \phi =0 \label{eq_ssl}
\end{equation}
Where $\eta_0$ is positive (since we assume $p(z)>0$). $\cos(k_e \tilde{z})$ is the periodicity supposed to create some periodic solutions inside the BG induced by  $\cos(2 \tilde{z})$(with $\beta=0$). Eq.~\eqref{eq_ssl} has been studied in the general case~\cite{Davis1980,Zounes1998}: actually, for some sets of parameters $\eta_0,\alpha$ and $\beta\ne 0$,  Eq.~\eqref{eq_ssl} has periodic solutions that would be unbounded if $\beta=0$. 
 
\begin{figure}
\begin{center}
\includegraphics[width=7cm]{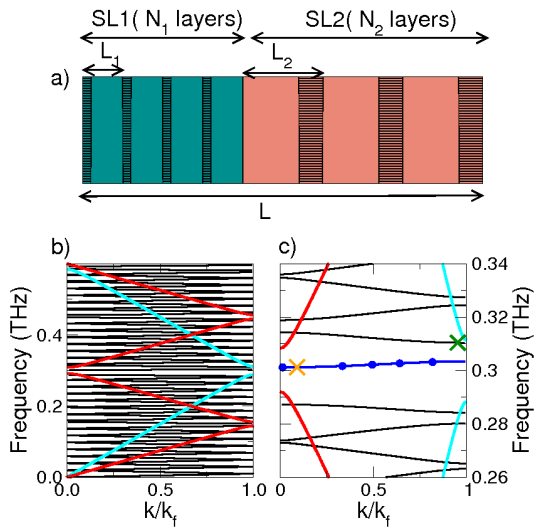}
\caption{a)Sketch of the elementary cell of a SSL. The cyan and red backgrounds correspond to the SL1 and SL2 regions, dashed regions underline the stratified structure of each SL. b) and c) Dispersion diagram of the SSL (black), SL1 (cyan) and SL2 (red)  between 0-0.6 THz and 0.26-0.34 THz,  the x-axis reports the Block wave vector normalized by $k_f =\frac{\pi}{\kappa}$ with $\kappa=L$(SSL)$,L_1$(SL1) or $L_2$(SL2).   In c), the blue point curve and orange and green crosses point up a SSL acoustic interface modes band and the SSL modes at   $\nu=0.3012$ THz (see Fig.~\ref{fig3}) and $\nu=0.3104$ THz. } 
\label{fig2}
\end{center}
\end{figure}

\begin{figure}
\begin{center}
\includegraphics[width=7cm]{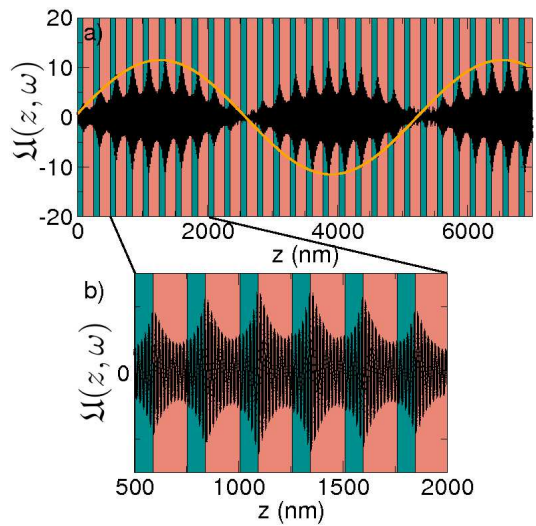}
\caption{Displacement fields $\mathfrak{U} (z,\omega)$ solutions of the  Eq.~\eqref{eq_wave2} in the SSL for $\nu=0.3012$ THz  between 0-7000nm a)  and 500-2000 nm b). The cyan and red background colors identifie the SL1 and SL2 regions. An orange curve, a guide to the eye emphasizes the oscillations at the Block wave vector. } 
\label{fig3}
\end{center}
\end{figure}

However, rather than discussing  on this theoretical equation, we propose to apply the previous concept to the propagation of phonons in an experimentally realizable structure. Such structure involves two periodicities:  we propose to use a SuperSuperLattice (SSL) (see Fig.~\ref{fig2}a): a $L$-periodic structure whose elementary unit cell is composed of two superlattices (SL) SL1 and SL2: respectively,  $10+x$ periods of a 5.65 nm (10 monolayers(ML))/2.26 nm (4ML) GaAs/AlAs SL (period $L_1$) with $x=0.5$ and  $10$ periods of a 11.3 nm (20ML)/4.52 nm (8ML) GaAs/AlAs SL  (period $L_2$)~\footnote{the (001) direction is perpendicular to the layers}.
Considering wave vectors with normal incidence, Fig.~\ref{fig2}b reports the dispersion diagrams of the SSL, SL1 and SL2  and, Fig.~\ref{fig2}c a zoom in the frequency range 0.26 THz-0.34 THz, corresponding to an overlapping region of SL1 and SL2 BGs. These dispersion diagrams are calculated from the transfer matrix~\cite{He1988}, a representation of the propagator of Eq.~\eqref{eq_wave2}  in these systems.  In Fig.~\ref{fig2}b, the multiple foldings and the mini-BGs created by the periodicity $L$ of the SSL appear.  More interestingly, the SSL phase diagram shows some phonons (blue point curve in Fig.~\ref{fig2} c) referred in the following as {\bf interface acoustic phonons} in the overlapping region of SL1 and SL2 BGs.  Fig.~\ref{fig3} reports the displacement fields $\mathfrak{U} (z,\omega)$, solutions of Eq.~\eqref{eq_wave2} in the SSL for  $\nu=0.3012$ THz, an interface acoustic phonons identified by an orange cross in Fig.~\ref{fig2}c).   Though in the BGs of both SL1 and SL2, this solution describes a periodic mode that thus propagate in an LUIP medium.  $\mathfrak{U} (z,\omega)$ is a oscillating function whose envelop is alternatively exponentially increasing and decreasing.  
The Fourier analysis of this mode evidences three characteristic lengths:  a short wavelength  (roughly $2 L_1=L_2$) oscillation exhibited in Fig.~\ref{fig3}b, an intermediate length $\approx L$, related to the exponential variation of the wave amplitude and, a long $ \gg L$ (orange curve in Fig.~\ref{fig3}a) characteristic length related to the Bloch wave vector of this mode reported in Fig.~\ref{fig2}c.    

These interface acoustic modes precisely correspond to some periodic solutions of the parametric oscillator in the phase space $\eta_0<0$ of Fig.~\ref{fig1}b). 
   We believe that the physical mechanism described here also leads to the existence of interface optic phonons~\cite{Sood1985} in SLs: the atomic potential, the fast periodicity (term $\cos(2 \tilde{z})$ in  Eq.~\eqref{eq_ssl}) induces the BG between acoustic and optic phonons; the SL periodicity (term $\cos(k_e \tilde{z})$ in  Eq.~\eqref{eq_ssl}) stands for the parametric stabilizing excitation. The common description of interface optic phonons involves a frequency dependent dielectric constant accounting for the fast periodicity~\cite{Camley1984}. 
   
Besides the interface acoustic modes and similarly to optic phonons, we also evidence in Fig.~\ref{fig2}  some {\bf confined acoustic phonons}. Their frequency belongs to only one of the BG of SL1 or SL2: for instance the mode at $\nu=0.3104 THz$ (green cross in Fig.~\ref{fig2}c),  in the BG of SL1 but ouside the BG of SL2. These confined acoustic phonons correspond to some periodic solutions of the parametric oscillator in the regions  $\eta_0<0$ or $\eta_0>0$ of Fig.~\ref{fig1}b). A detailed discussion will be reported elsewhere.

\begin{figure}
\begin{center}
\includegraphics[width=7cm]{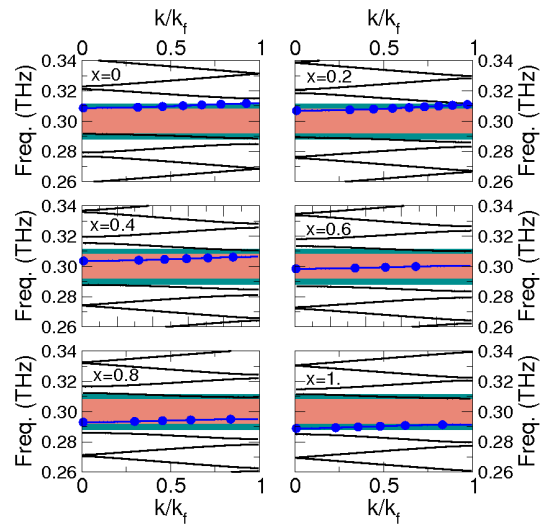}
\caption{Dispersion diagram of the SSL between 0.26-0.34 THz as a function of $x$. The BGs of SL1 and SL2 are represented by the cyan+red and red regions.  } 
\label{fig4}
\end{center}
\end{figure}

Beyond the existence of the interface acoustic modes, it is possible to fully control their frequencies from the engineering of the SSL, and more precisely by using fractional number of the periods of SL1 or SL2. We consider a SSL formed by  $10+x$ periods of  SL1 and  $10$ periods of SL2 and vary $x$ from $0$ to $1$. Fig.~\ref{fig4} shows a band frequency (blue point curve) in the dispersion diagram that continuously shifts from above to below the overlapping region of SL1 and SL2 BGs. Hence, by judiciously choosing the value of $x$, some interface or confined acoustic modes at arbitrarily frequencies can be produced. 

As already mentioned the concept described here is totally general and can formally  apply to any kind of waves. We now discuss this possibility for EM waves.  
 The propagation of an EM wave in a structure approaching the one of a SSL is related to the existence of optic Tamm state(OTS). 
 Kavokin et al.~\cite{Kavokin2005}  have considered the propagation of EM waves across two different
semi-infinite Photonic Crystals(PC) PC1 and PC2 with a common interface and with overlapping stop bands (or BGs). 
 By assuring the continuity of the in-plane components of the electric and magnetic fields at the interface between the two PCs, they evidence some modes, called OTS in the overlapping region of both PC1 and PC2 stop bands. OTS have been experimentally evidenced recently~\cite{goto2008}. 
 
Similarly to the work performed for phonons, we calculate(not shown) the phase diagram of a SuperPhotonic Crystal (SPC) whose unit cell is composed by (5+0.4) periods of a PC  $Ta_2O_5/ SiO_2$ 101.6 nm/149.2 nm  and 5 periods of a PC $Bi:YIG/ SiO_2$ 86 nm/138 nm~\cite{goto2008}~\footnote{the Phase diagram is calculated using the following optical indices; $n_{SiO_2}  = 2.07$, $n_{ Ta_2O_5}  = 4.41$ and $n_{Bi:YIG}  = 5.58$}.  We exhibit a band of periodic modes of energy $E \simeq 1.5eV$ in the overlapping region of BGs of both PCs.  The corresponding  modes, referred in the following as periodic OTS are similar to the interface acoustic phonons depicted in Fig.~\ref{fig3}. Moreover, as expected, the transmission spectra of a finite SPC (with several unit cells) exhibits transmission peaks corresponding to these modes. 

The concept described in this paper not only relates OTS to the physic of the parametric oscillator, but also extends the idea of Kavokin et al. showing the possibility to simultaneously assure the continuity conditions at all interfaces of an infinite succession of different finite PCs.

  As a conclusion, we would like to underline the intrinsic relation between the propagation of wave in LUIP media and the physics of the parametric oscillator.  
  Especially, we have shown how we can transpose the possibility to stabilize an inverse parametric pendulum to the case of waves. This concept is very general and can formally apply to any kind of waves: while it is well known that the periodicity induces some forbidden band frequency, a periodicity can also stabilize solutions that were expected to be unbounded.  We have chosen here to develop this concept to the case of elastic waves evidencing some new classes of phonons in realizable structures.  The case of EM waves has been discussed:  we have evidenced the relation between the inverse parametric pendulum and the OTS and have extended their existence  to periodic OTS.\\
  This work opens numerous possibilities of applications of this concept.  Due to the high localized amplitude of interface type modes,  exhibited devices (SSL or SPC) are expected to be useful in the investigation of non-linear effects or in the improvements of  electron-phonon or electron-photon couplings.  Moreover, they can also be used to elaborate advanced interferential filters.\\
   Finally, numerous perspectives to this work can be thought about: one can consider to apply the present concept to any kind of waves: electronic, spin, capillary waves\dots While our study has focused on normal incident waves in lossless materials, an extension to oblique incidence and  the investigation of the effect of absorption are required. Though  we have studied an unidimensional system, the generalization of the described concept  to 2 and 3-dimensional devices is worth being considered.  

\begin{acknowledgments}
The author thanks J. Morillo, M. Benoit,  A. Ponchet and J.R. Huntzinger  for useful discussions.   
\end{acknowledgments}


\begin{thebibliography}{3}%
\makeatletter
\providecommand \@ifxundefined [1]{%
 \@ifx{#1\undefined}
}%
\providecommand \@ifnum [1]{%
 \ifnum #1\expandafter \@firstoftwo
 \else \expandafter \@secondoftwo
 \fi
}%
\providecommand \@ifx [1]{%
 \ifx #1\expandafter \@firstoftwo
 \else \expandafter \@secondoftwo
 \fi
}%
\providecommand \natexlab [1]{#1}%
\providecommand \enquote  [1]{``#1''}%
\providecommand \bibnamefont  [1]{#1}%
\providecommand \bibfnamefont [1]{#1}%
\providecommand \citenamefont [1]{#1}%
\providecommand \href@noop [0]{\@secondoftwo}%
\providecommand \href [0]{\begingroup \@sanitize@url \@href}%
\providecommand \@href[1]{\@@startlink{#1}\@@href}%
\providecommand \@@href[1]{\endgroup#1\@@endlink}%
\providecommand \@sanitize@url [0]{\catcode `\\12\catcode `\$12\catcode
  `\&12\catcode `\#12\catcode `\^12\catcode `\_12\catcode `\%12\relax}%
\providecommand \@@startlink[1]{}%
\providecommand \@@endlink[0]{}%
\providecommand \url  [0]{\begingroup\@sanitize@url \@url }%
\providecommand \@url [1]{\endgroup\@href {#1}{\urlprefix }}%
\providecommand \urlprefix  [0]{URL }%
\providecommand \Eprint [0]{\href }%
\@ifxundefined \urlstyle {%
  \providecommand \doi  [0]{\begingroup \@sanitize@url \@doi}%
  \providecommand \@doi [1]{\endgroup \@@startlink {\doibase
  #1}doi:\discretionary {}{}{}#1\@@endlink }%
}{%
  \providecommand \doi  [0]{doi:\discretionary{}{}{}\begingroup
  \urlstyle{rm}\Url }%
}%
\providecommand \doibase [0]{http://dx.doi.org/}%
\providecommand \Doi [0]{\begingroup \@sanitize@url \@Doi }%
\providecommand \@Doi  [1]{\endgroup\@@startlink{\doibase#1}\@@Doi}%
\providecommand \@@Doi [1]{#1\@@endlink}%
\providecommand \selectlanguage [0]{\@gobble}%
\providecommand \bibinfo  [0]{\@secondoftwo}%
\providecommand \bibfield  [0]{\@secondoftwo}%
\providecommand \translation [1]{[#1]}%
\providecommand \BibitemOpen [0]{}%
\providecommand \bibitemStop [0]{}%
\providecommand \bibitemNoStop [0]{.\EOS\space}%
\providecommand \EOS [0]{\spacefactor3000\relax}%
\providecommand \BibitemShut  [1]{\csname bibitem#1\endcsname}%
\bibitem [{Note1()}]{Note1}%
  \BibitemOpen
  \bibinfo {note} {Actually, Hill equations solutions are qualitatively
  identical to Mathieu equation solutions.}\BibitemShut {Stop}%
\bibitem [{Note2()}]{Note2}%
  \BibitemOpen
  \bibinfo {note} {The (001) direction is perpendicular to the
  layers}\BibitemShut {NoStop}%
\bibitem [{Note3()}]{Note3}%
  \BibitemOpen
  \bibinfo {note} {The Phase diagram is calculated using the following optical
  indices; $n_{SiO_2} = 2.07$, $n_{ Ta_2O_5} = 4.41$ and $n_{Bi:YIG} =
  5.58$}\BibitemShut {NoStop}%
\end{thebibliography}%


\begin{thebibliography}{11}%
\makeatletter
\providecommand \@ifxundefined [1]{%
 \ifx #1\undefined \expandafter \@firstoftwo
 \else \expandafter \@secondoftwo
\fi
}%
\providecommand \@ifnum [1]{%
 \ifnum #1\expandafter \@firstoftwo
 \else \expandafter \@secondoftwo
\fi
}%
\providecommand \enquote [1]{``#1''}%
\providecommand \bibnamefont  [1]{#1}%
\providecommand \bibfnamefont [1]{#1}%
\providecommand \citenamefont [1]{#1}%
\providecommand\href[0]{\@sanitize\@href}%
\providecommand\@href[1]{\endgroup\@@startlink{#1}\endgroup\@@href}%
\providecommand\@@href[1]{#1\@@endlink}%
\providecommand \@sanitize [0]{\begingroup\catcode`\&12\catcode`\#12\relax}%
\@ifxundefined \pdfoutput {\@firstoftwo}{%
 \@ifnum{\z@=\pdfoutput}{\@firstoftwo}{\@secondoftwo}%
}{%
 \providecommand\@@startlink[1]{\leavevmode}%
 \providecommand\@@endlink[0]{}%
}{%
 \providecommand\@@startlink[1]{%
  \leavevmode
  \pdfstartlink
   attr{/Border[0 0 1 ]/H/I/C[0 1 1]}%
   user{/Subtype/Link/A<</Type/Action/S/URI/URI(#1)>>}%
  \relax
 }%
 \providecommand\@@endlink[0]{\pdfendlink}%
}%
\providecommand \url  [0]{\begingroup\@sanitize \@url }%
\providecommand \@url [1]{\endgroup\@href {#1}{\urlprefix}}%
\providecommand \urlprefix [0]{URL }%
\providecommand \Eprint[0]{\href }%
\@ifxundefined \urlstyle {%
  \providecommand \doi [1]{doi:\discretionary{}{}{}#1}%
}{%
  \providecommand \doi [0]{doi:\discretionary{}{}{}\begingroup
  \urlstyle{rm}\Url }%
}%
\providecommand \doibase [0]{http://dx.doi.org/}%
\providecommand \Doi[1]{\href{\doibase#1}}%
\providecommand \bibAnnote [3]{%
  \BibitemShut{#1}%
  \begin{quotation}\noindent
    \textsc{Key:}\ #2\\\textsc{Annotation:}\ #3%
  \end{quotation}%
}%
\providecommand \bibAnnoteFile [2]{%
  \IfFileExists{#2}{\bibAnnote {#1} {#2} {\input{#2}}}{}%
}%
\providecommand \typeout [0]{\immediate \write \m@ne }%
\providecommand \selectlanguage [0]{\@gobble}%
\providecommand \bibinfo [0]{\@secondoftwo}%
\providecommand \bibfield [0]{\@secondoftwo}%
\providecommand \translation [1]{[#1]}%
\providecommand \BibitemOpen[0]{}%
\providecommand \bibitemStop [0]{}%
\providecommand \bibitemNoStop [0]{.\EOS\space}%
\providecommand \EOS [0]{\spacefactor3000\relax}%
\providecommand \BibitemShut [1]{\csname bibitem#1\endcsname}%
\bibitem{Istrate2006}%
  \BibitemOpen
  \bibfield{author}{%
  \bibinfo {author} {\bibfnamefont{E.}~\bibnamefont{Istrate}}\ and\ \bibinfo
  {author} {\bibfnamefont{E.~H.}\ \bibnamefont{Sargent}},\ }%
  \bibfield{journal}{%
  \Doi{10.1103/RevModPhys.78.455}{\bibinfo {journal} {Rev. Mod. Phys.}}\ }%
  \textbf{\bibinfo {volume} {78}},\ \bibinfo {pages} {455} (\bibinfo {year}
  {2006})%
  \bibAnnoteFile{NoStop}{Istrate2006}%
\bibitem{Boudouti2009}%
  \BibitemOpen
  \bibfield{author}{%
  \bibinfo {author} {\bibfnamefont{E.~E.}\ \bibnamefont{Boudouti}}, \bibinfo
  {author} {\bibfnamefont{et al.}}, \ }%
  \bibfield{journal}{%
  \Doi{10.1016/j.surfrep.2009.07.005}{\bibinfo {journal} {Surf. Sci.
  Rep.}}\ }%
  \textbf{\bibinfo {volume} {64}},\ \bibinfo {pages} {471 } (\bibinfo {year}
  {2009})%
  \bibAnnoteFile{NoStop}{Boudouti2009}%
\bibitem{ashcroft}%
  \BibitemOpen
  \bibfield{author}{%
  \bibinfo {author} {\bibfnamefont{N.}~\bibnamefont{Ashcroft}}\ and\ \bibinfo
  {author} {\bibfnamefont{N.~D.}\ \bibnamefont{Mermin}},\ }%
  \emph{\bibinfo {title} {Solid State Physics}}\ (\bibinfo {publisher} {Holt,
  Rinehart and Winston},\ \bibinfo {year} {1976})%
  \bibAnnoteFile{NoStop}{ashcroft}%
\bibitem{Steslicka2002}%
  \BibitemOpen
  \bibfield{author}{%
  \bibinfo {author} {\bibfnamefont{M.}~\bibnamefont{Steslicka}}, \bibinfo
  {author} {\bibfnamefont{et al.}},\ }%
  \bibfield{journal}{%
  \Doi{doi:10.1016/S0167-5729(02)00052-3}{\bibinfo {journal} {Surf. Sci.
  Rep.}}\ }%
  \textbf{\bibinfo {volume} {47}},\ \bibinfo {pages} {92} (\bibinfo {year}
  {2002})%
  \bibAnnoteFile{NoStop}{Steslicka2002}%
\bibitem{Combe2009a}%
  \BibitemOpen
  \bibfield{author}{%
  \bibinfo {author} {\bibfnamefont{N.}~\bibnamefont{Combe}}, \bibinfo {author}
  {\bibfnamefont{J.~R.}\ \bibnamefont{Huntzinger}},\ and\ \bibinfo {author}
  {\bibfnamefont{J.}~\bibnamefont{Morillo}},\ }%
  \bibfield{journal}{%
  \Doi{10.1140/epjb/e2009-00061-3}{\bibinfo {journal} {Euro. Phys. J. B}}\ }%
  \textbf{\bibinfo {volume} {68}},\ \bibinfo {pages} {47} (\bibinfo {year}
  {2009})%
  \bibAnnoteFile{NoStop}{Combe2009a}%
\bibitem{Kapitza1965}%
  \BibitemOpen
  \bibfield{author}{%
  \bibinfo {author} {\bibfnamefont{P.}~\bibnamefont{Kapitza}},\ }%
  in\ \emph{\bibinfo {booktitle} {Collected Papers by P.L. Kapitza}},\
  Vol.~\bibinfo {volume} {2},\ \bibinfo {editor} {edited by\ \bibinfo {editor}
  {\bibfnamefont{D.~T.}\ \bibnamefont{Haar}}}\ (\bibinfo {publisher} {Pergamon
  Press},\ \bibinfo {year} {1965})\ pp.\ \bibinfo {pages} {714--726}%
  \bibAnnoteFile{NoStop}{Kapitza1965}%
\bibitem{Note1}%
  \BibitemOpen
  \bibinfo {note} {Actually, Hill equations solutions are qualitatively
  identical to Mathieu equation solutions.}%
  \bibAnnoteFile{Stop}{Note1}%
\bibitem{Palik1997}%
  \BibitemOpen
  \bibfield{author}{%
  \bibinfo {author} {\bibfnamefont{E.~D.}\ \bibnamefont{Palik}},\ }%
  \emph{\bibinfo {title} {Handbook of optical constants of solids}}\ (\bibinfo
  {publisher} {Academic Press},\ \bibinfo {year} {1997})%
  \bibAnnoteFile{NoStop}{Palik1997}%
\bibitem{Davis1980}%
  \BibitemOpen
  \bibfield{author}{%
  \bibinfo {author} {\bibfnamefont{S.~H.}\ \bibnamefont{Davis}}\ and\ \bibinfo
  {author} {\bibfnamefont{S.}~\bibnamefont{Rosenblat}},\ }%
  \bibfield{journal}{%
  \Doi{10.1137/0138012}{\bibinfo {journal} {SIAM J. Appl.
  Math.}}\ }%
  \textbf{\bibinfo {volume} {38}},\ \bibinfo {pages} {139} (\bibinfo {year}
  {1980})%
  \bibAnnoteFile{NoStop}{Davis1980}%
\bibitem{Zounes1998}%
  \BibitemOpen
  \bibfield{author}{%
  \bibinfo {author} {\bibfnamefont{R.~S.}\ \bibnamefont{Zounes}}\ and\ \bibinfo
  {author} {\bibfnamefont{R.~H.}\ \bibnamefont{Rand}},\ }%
  \bibfield{journal}{%
  \Doi{10.1137/S0036139996303877}{\bibinfo {journal} {SIAM J. Appl. Math.}}\ }%
  \textbf{\bibinfo {volume} {58}},\ \bibinfo {pages} {1094} (\bibinfo {year}
  {1998})%
  \bibAnnoteFile{NoStop}{Zounes1998}%
\bibitem{Note2}%
  \BibitemOpen
  \bibinfo {note} {The (001) direction is perpendicular to the layers}%
  \bibAnnoteFile{NoStop}{Note2}%
\bibitem{He1988}%
  \BibitemOpen
  \bibfield{author}{%
  \bibinfo {author} {\bibfnamefont{J.}~\bibnamefont{He}}, \bibinfo {author}
  {\bibfnamefont{B.}~\bibnamefont{Djafari-Rouhani}},\ and\ \bibinfo {author}
  {\bibfnamefont{J.}~\bibnamefont{Sapriel}},\ }%
  \bibfield{journal}{%
  \Doi{10.1103/PhysRevB.37.4086}{\bibinfo {journal} {Phys. Rev. B}}\ }%
  \textbf{\bibinfo {volume} {37}},\ \bibinfo {pages} {4086} (\bibinfo {year}
  {1988})%
  \bibAnnoteFile{NoStop}{He1988}%
\bibitem{Sood1985}%
  \BibitemOpen
  \bibfield{author}{%
  \bibinfo {author} {\bibfnamefont{A.~K.}\ \bibnamefont{Sood}}, \bibinfo
  {author} {\bibfnamefont{et al.}}, \ }%
  \bibfield{journal}{%
  \Doi{10.1103/PhysRevLett.54.2115}{\bibinfo {journal} {Phys. Rev. Lett.}}\ }%
  \textbf{\bibinfo {volume} {54}},\ \bibinfo {pages} {2115} (\bibinfo {year}
  {1985})%
  \bibAnnoteFile{NoStop}{Sood1985}%
\bibitem{Camley1984}%
  \BibitemOpen
  \bibfield{author}{%
  \bibinfo {author} {\bibfnamefont{R.~E.}\ \bibnamefont{Camley}}\ and\ \bibinfo
  {author} {\bibfnamefont{D.~L.}\ \bibnamefont{Mills}},\ }%
  \bibfield{journal}{%
  \Doi{10.1103/PhysRevB.29.1695}{\bibinfo {journal} {Phys. Rev. B}}\ }%
  \textbf{\bibinfo {volume} {29}},\ \bibinfo {pages} {1695} (\bibinfo {year}
  {1984})%
  \bibAnnoteFile{NoStop}{Camley1984}%
\bibitem{Kavokin2005}%
  \BibitemOpen
  \bibfield{author}{%
  \bibinfo {author} {\bibfnamefont{A.~V.}\ \bibnamefont{Kavokin}}, \bibinfo
  {author} {\bibfnamefont{I.~A.}\ \bibnamefont{Shelykh}},\ and\ \bibinfo
  {author} {\bibfnamefont{G.}~\bibnamefont{Malpuech}},\ }%
  \bibfield{journal}{%
  \Doi{10.1103/PhysRevB.72.233102}{\bibinfo {journal} {Phys. Rev. B}}\ }%
  \textbf{\bibinfo {volume} {72}},\ \bibinfo {pages} {233102} (\bibinfo {year}
  {2005})%
  \bibAnnoteFile{NoStop}{Kavokin2005}%
\bibitem{goto2008}%
  \BibitemOpen
  \bibfield{author}{%
  \bibinfo {author} {\bibfnamefont{T.}~\bibnamefont{Goto}}, \bibinfo {author}
  {\bibfnamefont{et al.}},\ }%
  \bibfield{journal}{%
  \Doi{10.1103/PhysRevLett.101.113902}{\bibinfo {journal} {Phys. Rev. Lett.}}\
  }%
  \textbf{\bibinfo {volume} {101}},\ \bibinfo {pages} {113902} (\bibinfo {year} {2008})%
  \bibAnnoteFile{NoStop}{goto2008}%
  \bibitem{Note3}%
  \BibitemOpen
  \bibinfo {note} {Optical indices used are; $n_{SiO_2}  = 2.07$, $n_{ Ta_2O_5}  = 4.41$ and $n_{Bi:YIG}  = 5.58$}%
  \bibAnnoteFile{NoStop}{Note3}%
\end{thebibliography}

%

\end{document}